\documentstyle[multicol,aps,epsf]{revtex}
\begin{document}
\title{Ballistic Coalescence Model}

\author{S.~Ispolatov and P.~L.~Krapivsky}
\address{Center for Polymer Studies and Department of Physics,
Boston University, Boston, MA 02215}

\maketitle 
\begin{abstract}

We study statistical properties of a one dimensional infinite system of
coalescing particles.  Each particle moves with constant velocity $\pm
v$ towards its closest neighbor and merges with it upon collision.  We
propose a mean-field theory that confirms a $t^{-1}$ concentration decay
obtained in simulations and provides qualitative description for the
densities of growing, constant, and shrinking inter-particle gaps.

\end{abstract}
\begin{multicols}{2}

We introduce a simple deterministic model describing a coarsening
dynamics of interacting domains.  We arrived at this model in attempt to
model the essential features of development of countries and
civilizations.  In our model, the civilizations are represented by
domains that continuously cover one-dimensional ``world'' without gaps
and overlaps.  Neighboring civilizations are engaged in a permanent
warfare: A bigger civilization invades the lesser one, so that the
border (interface) moves with velocity $\pm v/2$ with $v$ being the same
for all pairs of neighbors.  The model is equivalent to an infinite
particle aggregating system in which each particle (corresponding to the
interface between domains) moves ballistically towards its closest
neighbor.  When a civilization shrinks to zero size and disappears, two
particles, corresponding to its borders, collide.  Such collisions
between interfaces lead to coalescence, $I+I\to I$ where $I$ symbolizes
an interface, and the emerging interface starts moving towards its
nearest neighbor with velocity of the same magnitude $v/2$ (Fig.~1).

\begin{figure}
\centerline{\epsfxsize=8cm \epsfbox{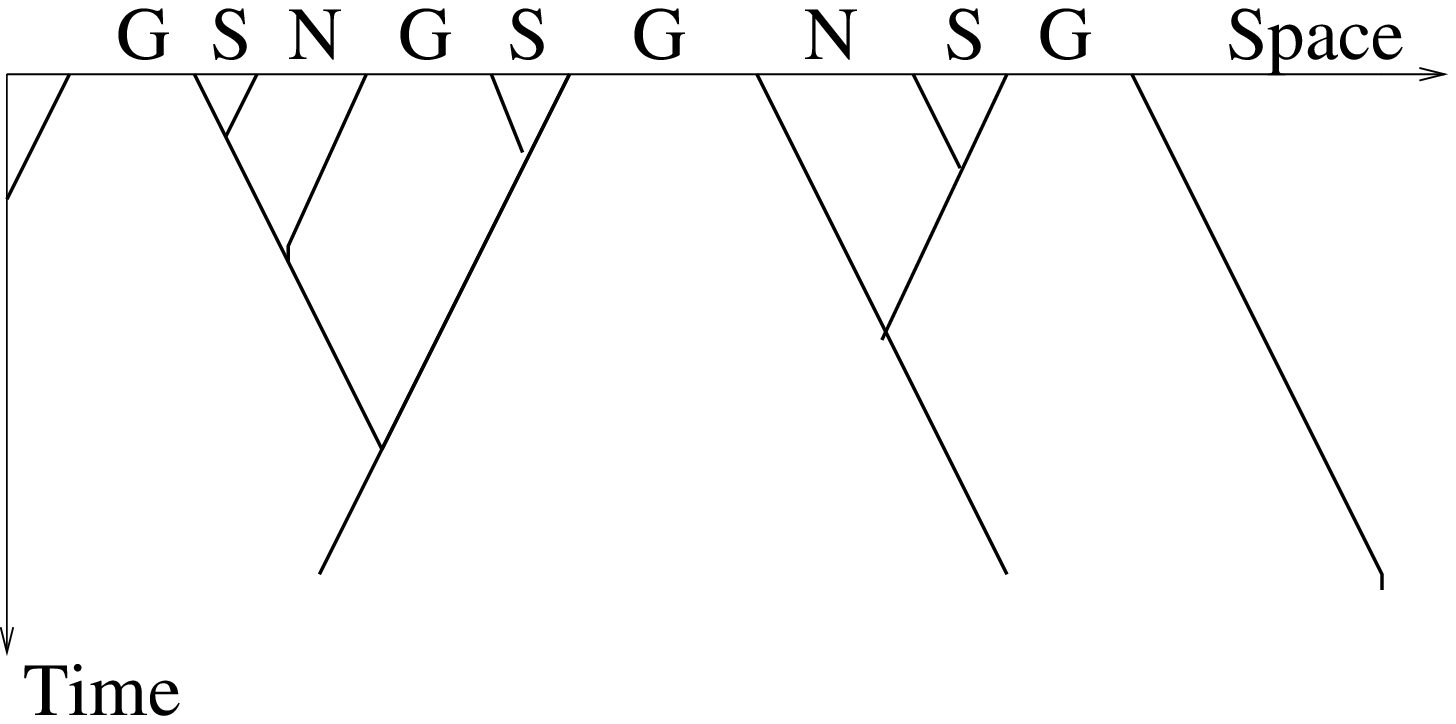}}
\noindent
{\small {\bf Fig.~1}. Schematic illustration of domain structure.  
Domains that grow, remain constant, and shrink at early time are 
denoted by G, N, and S, respectively.}
\label{Fig-def}
\end{figure}

The gradual character and general nature of the model makes questionable
its ability to reproduce specific details of human history that have
been actually observed.  However, it does provide some statistical
characteristics like size distribution and predicts decrease of the
total number of civilizations.  An appealing simplicity of our model
suggests that it might be relevant to description of other coarsening
phenomena.

In our previous work \cite{sps}, we studied a model with domains growing
freely in otherwise empty space and engaging in instantaneous warfare
upon contact.  One can view the model of Ref.~\cite{sps} and the present
model as two limiting cases of the general growth-and-war process: The
former {\em growth-controlled} process is limited by the growth rate
while the latter process is {\em war-controlled}.  Another model which
has strong relevance to our system is the ``cut-in-two'' model of
Ref.~\cite{d}, introduced in the context of breath figures
coarsening. Similarly to our model, it assumes the complete coverage
without overlaps, but unlike our model, coarsening events are
instantaneous.  Namely, evolution proceeds by consecutive elimination of
the current shortest domain.

In our model, which can be also called the ballistic coalescence model,
there are three types of domains -- growing, neutral (with constant
length), and shrinking. A growing domain is bigger than both of its
neighbors, so its length grows with velocity $v$.  A neutral domain has
one neighbor which is bigger and the other which is smaller, so its
length remains constant though its location does change.  Finally, a
shrinking domain is smaller than both of its neighbors, and its length
decreases in time with rate $v$.  When such shrinking domain disappears,
a change takes place in one of the two neighboring domains -- a growing
domain may become neutral, or a neutral domain may turn into a shrinking
one.  The number density of growing domains, $G(t)$, is always equal the
number density of shrinking domains, $S(t)$, since the space remains
continuously covered without gaps and overlaps throughout the process.
A growing domain of initial length $\ell_0$ has the length $vt+\ell_0$
at time $t$, implying that $G(t)\sim t^{-1}$ and $S(t)\sim t^{-1}$.  One
might expect that the number density of neutral domains, $N(t)$, decays
{\em slower} than $t^{-1}$, implying that the total domain density also
exhibits anomalously slow decay.  This indeed happens e.g. in apparently
similar model of ballistic annihilation, $I+I\to 0$, where $t^{-1/2}$
decay of the domain density is observed\cite{ef,esp}.  In both ballistic
annihilation and ballistic coalescence models, growing domains cannot be
neighbors, while an arbitrary number of consecutive neutral domains may
coexist.  However, in the ballistic annihilation model such a train of
neutral domains has a good chance to live long while in our model this
sequence of neutral domains can be eliminated by a single growing
domain.  Because of this relative vulnerability of neutral domains in
the ballistic coalescence model, their number density scales similarly
to the other two densities, $G(t)$ and $S(t)$.  Moreover, we will show
below that $N(t)$ becomes smaller than $G(t)$ and $S(t)$ as the process
develops.  Note that initially all the densities are the same,
$G(0)=N(0)=S(0)$.  Indeed, the initial sizes of domains are
uncorrelated, so if we take three consecutive intervals the
probabilities that the middle interval is the smallest, medium, or
biggest are all the same.

Thus in the long time limit the densities behave according to

\begin{equation}
\label{GNS}
G(t)={g\over t}, \quad
N(t)={n\over t}, \quad
S(t)={s\over t}, 
\end{equation}
with constants $g=s$ and $n$ to be determined.  Consider an arbitrary
shrinking domain of vanishing length.  It may be surrounded by two
neutral, two growing, or growing and neutral domains.  Denote by
$\lambda$, $\mu$, and $\nu$ the respective probabilities
($\lambda+\mu+\nu=1$).  The rate at which domains disappear, $vS(0,t)$,
may be written as $\alpha g/t^2$ with some constant $\alpha$.  Here
$S(x,t)$ is the density of shrinking domains of length $x$.  The
normalization $\int_0^t dx\,S(x,t)=S(t)\sim t^{-1}$ suggests $S(0,t)\sim
t^{-2}$ and explains the rate given above.  Now it is straightforward to
write down the rate equations

\begin{eqnarray}
\label{gns}
{d\over dt}\,{g\over t}&=&-{\alpha g\over t^2}\,\mu,\nonumber\\
{d\over dt}\,{n\over t}&=&-{\alpha g\over t^2}\,(-\mu+\nu+\lambda),\\
{d\over dt}\,{s\over t}&=&-{\alpha g\over t^2}\,(1-\nu-\lambda).\nonumber
\end{eqnarray}
Summing Eqs.~(\ref{gns}) gives $\alpha=(g+n+s)/g=2+n/g$.  The above
equations are exact.  To proceed further we need to know $\lambda, \mu,
\nu$. We {\em assume} that the types of domains surrounding vanishing
shrinking ones are uncorrelated, so the probability of picking up
growing or neutral domain is proportional to its respective
concentration.  Mathematically, it means that

\begin{equation}
\label{lmn}
\lambda=\left({n\over g+n}\right)^2, \quad
\mu=\left({g\over g+n}\right)^2, \quad
\nu={2gn\over (g+n)^2}. 
\end{equation}
Eqs.~(\ref{gns}) imply $\alpha\mu=1$ and $\alpha(1-2\mu)=n/g$.
Substituting $\alpha=2+n/g$ and $\mu=g^2(g+n)^{-2}$ into any of these
equations allows us to find $n/g$,

\begin{equation}
\label{golden}
{n\over g}={\sqrt{5}-1\over 2}\equiv \rho, 
\end{equation}
where $\rho$ is known as the ``golden ratio''.  We still need to
determine one constant, say $g$.  This can be accomplished by using
the fact that the system is continuously covered with domains.  We
need to determine the length distributions $G(x,t)$, $N(x,t)$, and
$S(x,t)$.  In the limit of large time, we may ignore differences in
sizes of growing domains.  These differences are determined by initial
size distribution and do not change with time.  Thus the length
distribution of growing domains is simply

\begin{equation}
\label{g}
G(x,t)={g\over t}\,\delta(x-t).
\end{equation}
Two other length distributions are expected to scale, so we write

\begin{equation}
\label{ns}
N(x,t)={g\over t^2}\,A\left({x\over t}\right), \quad
S(x,t)={g\over t^2}\,B\left({x\over t}\right).
\end{equation}

In the mean-field approximation (Eq.~(\ref{lmn})), the governing
equations for $N(x,t)$ and $S(x,t)$ read

\begin{equation}
\label{n}
{\partial N(x,t)\over \partial t}=-{\alpha g\over t^2}\,{N(x,t)\over N(t)}
\left[\nu
+2\lambda\,{\int_x^t dx'\,N(x',t)\over N(t)}\right],
\end{equation}

\begin{equation}
\label{s}
{\partial S\over \partial t}-{\partial S\over \partial x}
={\alpha g\over t^2}\,{N(x,t)\over N(t)}
\left[\nu+2\lambda\,{\int_x^t dx'\,N(x',t)\over N(t)}\right].
\end{equation}
The spatial derivative term in (\ref{s}) accounts for continuous
shrinking.  The velocity of shrinking $v$ is set equal to one without
loss of generality.  We also have dropped two terms proportional to
$\delta$-functions -- the gain term in Eq.~(\ref{n}) relevant for
conversion of growing domains into neutral, and the loss term in
Eq.~(\ref{s}) accounting for removal of zero-size shrinking domains.
These terms provide appropriate boundary conditions.

Using the scaling form (\ref{ns}) and relations (\ref{lmn}) we
simplify (\ref{n}) and (\ref{s}) to

\begin{equation}
\label{A}
uA'=2A\int_u^1 dv A(v)
\end{equation}
and
\begin{equation}
\label{B}
2B+(1+u)B'=-2A-2A\int_u^1 dv A(v).
\end{equation}
In these equations $u=x/t$, $A'=dA/du$, $B'=dB/du$.  According
to the definitions (\ref{GNS}) and (\ref{ns}), the  
scaling functions $A(u)$ and $B(u)$ obey normalization conditions

\begin{equation}
\label{norm}
\int_0^1 du\,A(u)=\rho, \quad
\int_0^1 du\,B(u)=1.
\end{equation}
Additionally, creation of the longest neutral domains ($x=t$) and
removal of the  zero-size shrinking domains give
$N(t,t)=\alpha\mu g/t^2$ and $S(0,t)=\alpha g/t^2$. This provides the
boundary conditions:

\begin{equation}
\label{10}
A(1)=1, \quad
B(0)=\alpha.
\end{equation}
Solving (\ref{A}) gives

\begin{equation}
\label{asol}
A(u)={(2+\rho^{-1})^2 u^{2\rho}\over (1+\rho^{-1}+u^{2\rho+1})^2}.
\end{equation}
Note that $A(0)=0$, in agreement with intuitive expectation that there
are no neutral domains of length zero.

To solve for the length distribution of shrinking domains we first
multiply (\ref{s}) on $1+u$ so that the left-hand side becomes a
complete derivative, $[(1+u)^2B]'$.  The right-hand side contains 
already known functions.  Integrating the resulting equation with the
boundary condition $B(0)=\alpha=2+\rho$ we obtain

\begin{eqnarray}
\label{bsol}
B(u)&=&{2+\rho\over (1+u)^2}-{1+2\rho\over (1+u)^2}\,
{u^{2\rho+1}\over 1+\rho^{-1}+u^{2\rho+1}}\\
&-&{\rho(1+\rho)(2+\rho^{-1})^2\over 1+u}\,
{u^{2\rho+1}\over (1+\rho^{-1}+u^{2\rho+1})^2}.\nonumber
\end{eqnarray}
One can check that $B(1)=0$, the result that agrees with intuitive
expectation that there are no shrinking domains of maximal possible
length $x=t$.  We verified that the normalization condition of
Eq.~(\ref{norm}) is satisfied.  The scaled length distributions of
neutral and shrinking domains are plotted on Figs.~2 and 3.

Now we can compute the constant $g$.  The requirement that the space
is completely covered,

\begin{equation}
\label{req}
\int_0^t dx\,x[G(x,t)+N(x,t)+S(x,t)]=1,
\end{equation}
gives

\begin{equation}
\label{cov}
g=\left(1+ \int_0^1 du\,u[A(u)+B(u)]\right)^{-1}
\cong 0.625837.
\end{equation}

We performed molecular dynamic simulations for systems of $2\cdot 10^6$
particles with periodic boundary conditions.  Results were averaged over
500 initial configurations.  We found that the ratio $n/g$ of neutral to
growing particle densities decreases slowly from 1 to approximately
0.66.  This value is $7\%$ larger than the theoretical prediction
$n/g=\rho\cong 0.618033989$; the statistical error for the measured
value is less than $3\%$.

\begin{figure}
\centerline{\epsfxsize=8cm \epsfbox{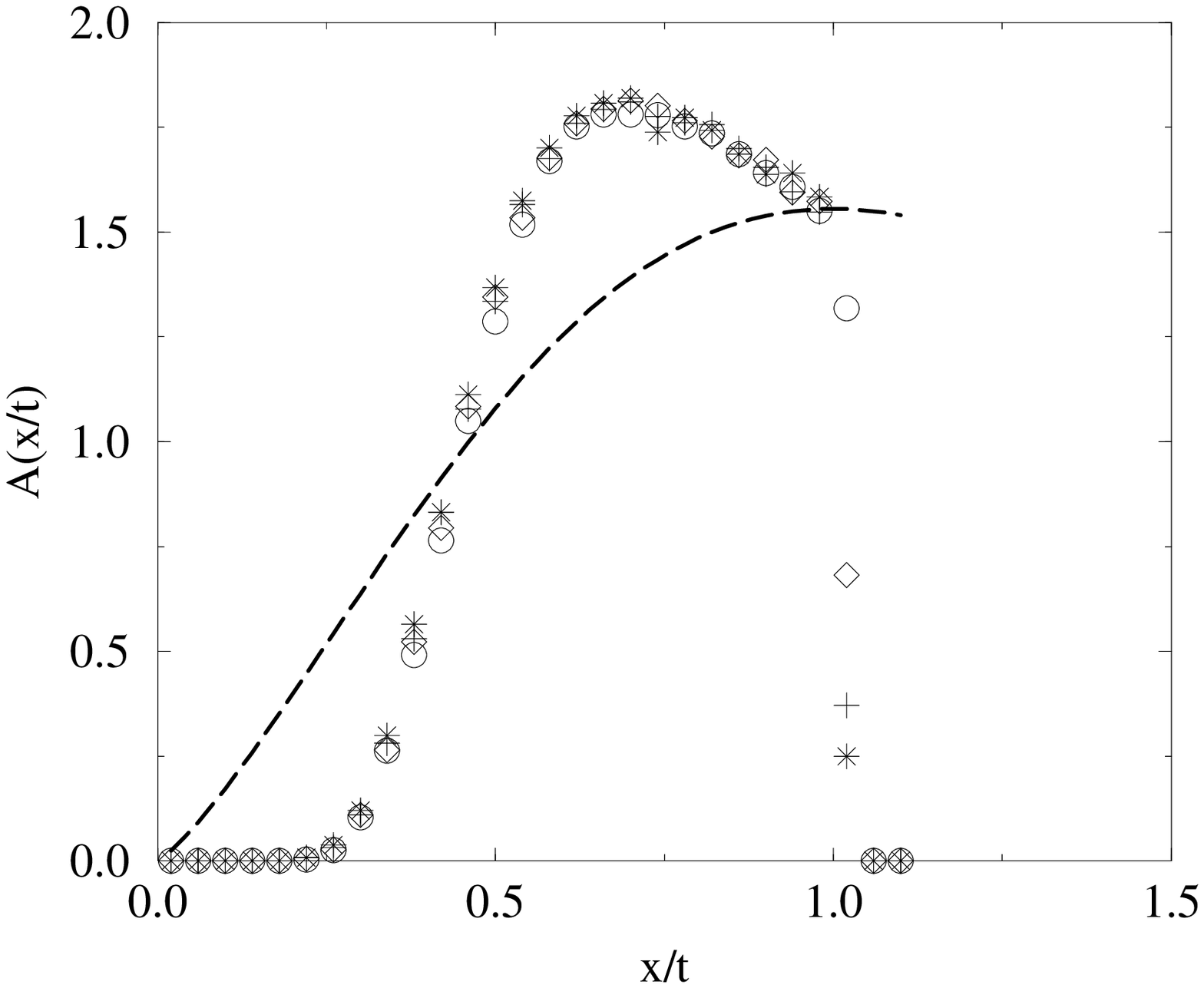}}
{\small {\bf Fig.~2}. Plot of scaled domain size distribution for 
neutral domains, $A(x/t)$.
Time is incremented by 2.25: $t=2192 (\bigcirc)$, $t=4932 
(\diamond)$, $t=11097 (+)$, $t=24968 (\star)$. Mean-field
prediction (Eq.~13) is shown by dashed line.}
\label{Fig-sim1}
\end{figure}

We also determined numerically the
probabilities $\lambda, \mu, \nu$ and found more pronounced difference
with mean-field predictions of Eq.~(\ref{lmn}): $\lambda\approx 0.1,
\mu\approx 0.5, \nu\approx 0.4$ vs. $\lambda_{\rm MFT}\cong
0.145898034,\mu_{\rm MFT}\cong 0.381966011, \nu_{\rm MFT}\cong
0.472135955$.  We found that scaling (\ref{ns}) for the neutral and
shrinking domain length distribution functions holds indeed, although
the shape of the experimental curves is different from mean-field
predictions (Figs.~2 and 3).  The discrepancy is particularly drastic
for the size distribution of neutral domains indicating that
correlations, not accounted by the mean-field theory, are especially
important for them.

\begin{figure}
\centerline{\epsfxsize=8cm \epsfbox{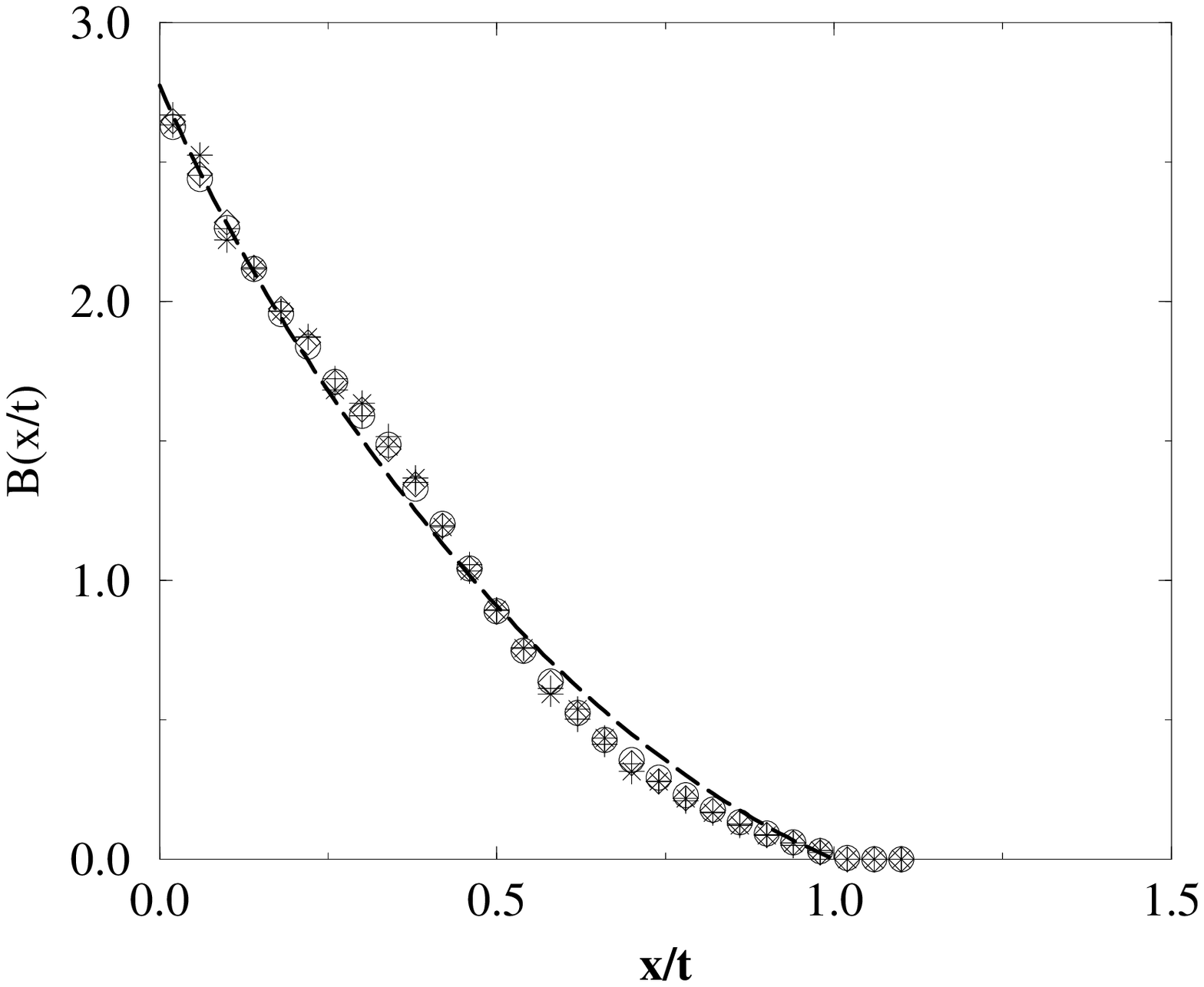}}
{\small {\bf Fig.~3}. Plot of scaled domain size distribution 
for shrinking domains, $B(x/t)$.
Time is incremented by 2.25: $t=2192 (\bigcirc)$, $t=4932 
(\diamond)$, $t=11097 (+)$, $t=24968 (\star)$. Mean-field
prediction (Eq.~14) is shown by dashed line.}
\label{Fig-sim2}
\end{figure}

\begin{figure}
\centerline{\epsfxsize=8cm \epsfbox{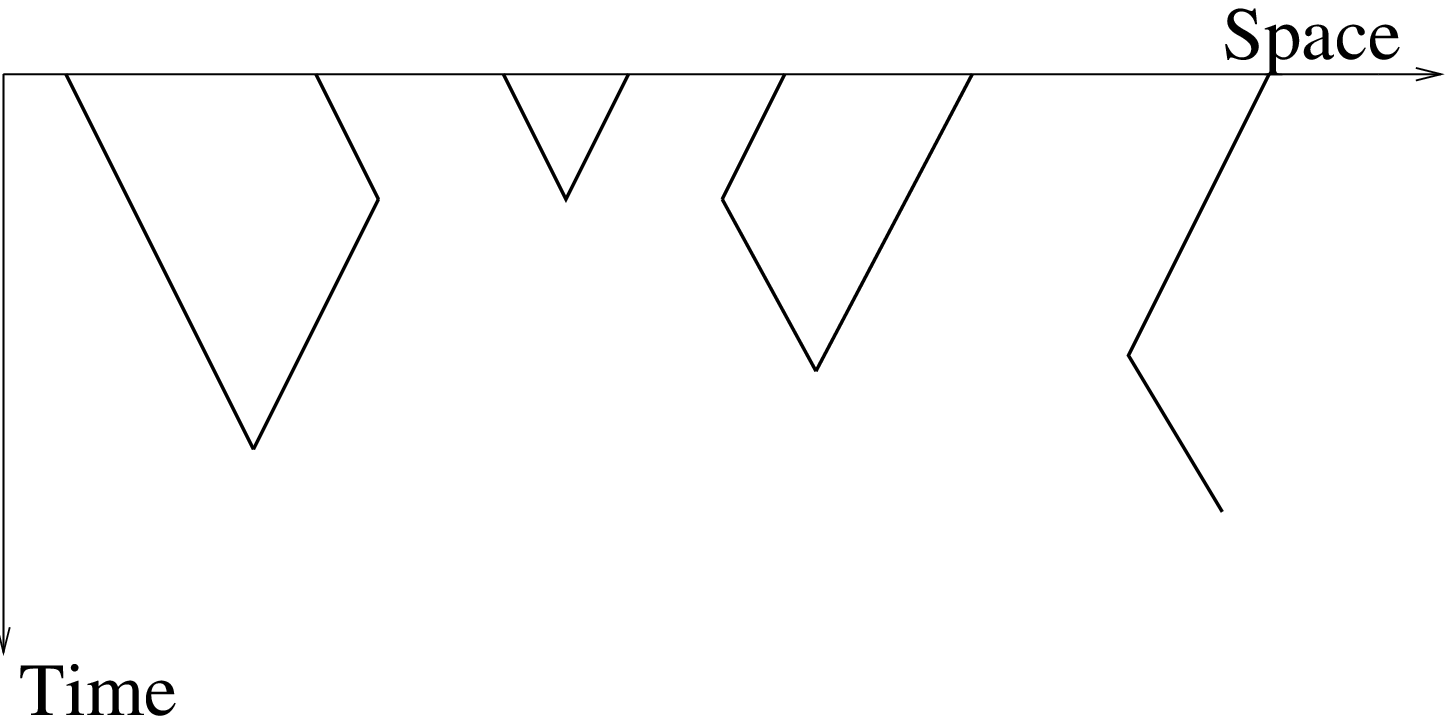}}
{\small {\bf Fig.~4}. Schematic illustration of domain structure
for a system with interface annihilation}
\label{Fig-def2}
\end{figure}

A natural generalization of this system is a model of an ``alternating
bipolar world''. Here neighboring civilizations are hostile to each
other so they are engaged in warfare, while collisions of friendly
civilizations lead to merge. In other words, interfaces annihilate upon
contact rather than coalesce as illustrated on Fig.~4.  When a shrinking
domain disappears, its left-hand side and right-hand side neighbors
merge into a single domain.  Since a domain length can instantaneously
increase upon each shrinking event, such events often lead to changes in
as many as four domains surrounding the shrinking one (central group of
domains on Fig.~4).  The model is thus different from the ballistic
annihilation model\cite{ef}.  Furthermore, correlations become more
important than in our original ballistic coalescence model.  However,
this interface annihilation model exhibits the same scaling as the
original interface coalescence model, {\it i.e.}, the length of a
typical growing, neutral, or shrinking domain grows linearly in time,
and domain concentration decays as $1/t$.  Scaled plots of length
distribution for growing, neutral, and shrinking domains are presented
of Fig.~5.  Note that unlike the original model, the length distribution
of growing domain remains unsingular and unbounded.

\begin{figure}
\centerline{\epsfxsize=8cm \epsfbox{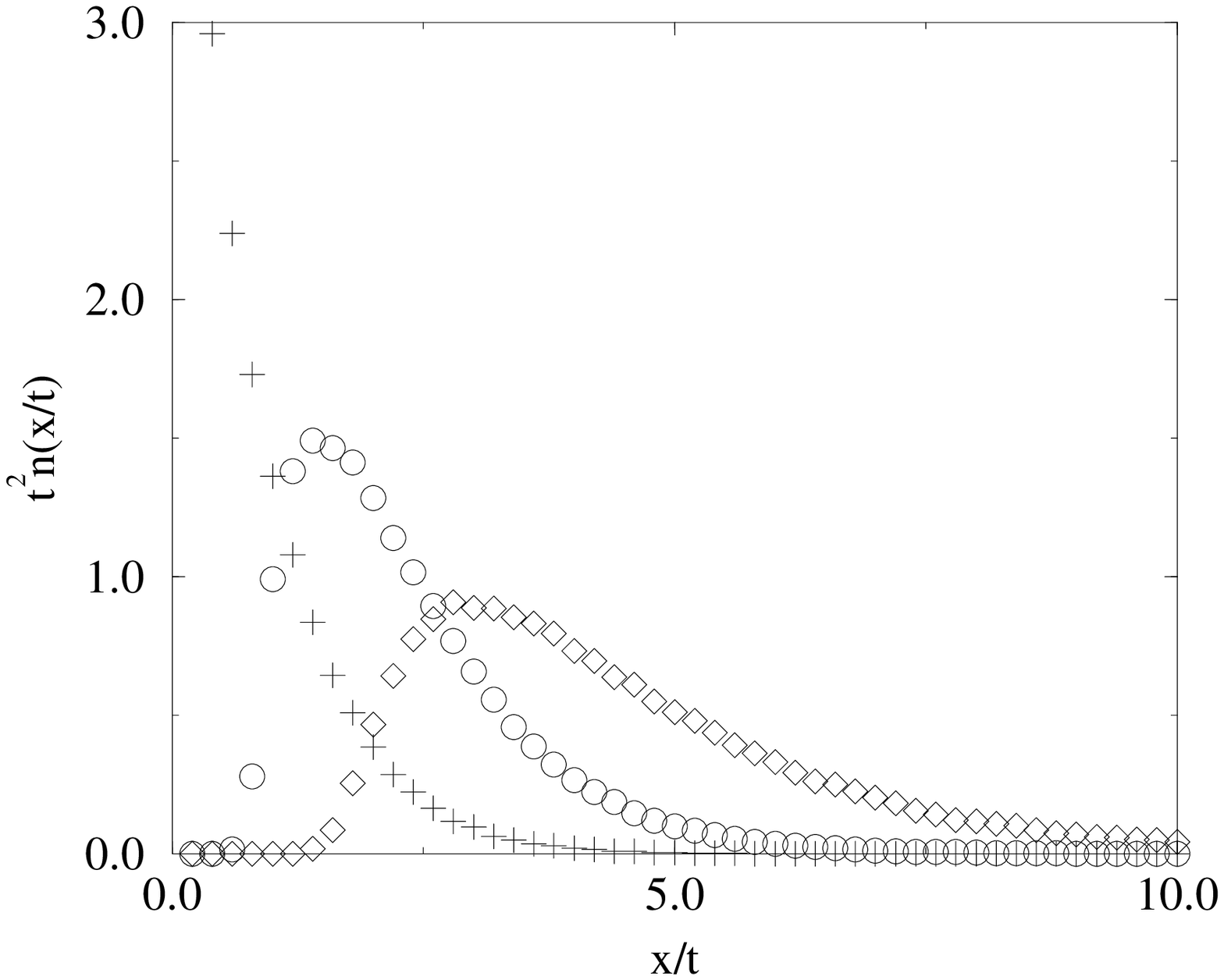}}
{\small {\bf Fig.~5}. Plots of scaled domain size distribution for
growing domains ($\diamond$), neutral domains ($\bigcirc$),
and shrinking domains ($+$) for interface annihilation model. 
Simulation included averaging over 500 configurations
of 2 million initial domains each.}
\label{Fig-sim3}
\end{figure}

Coarsening rules in the above models are similar to those in
one-dimensional Potts models with zero-temperature Glauber dynamics.
Indeed, in the infinite-state Potts model interfaces aggregate upon
collisions while in the 2-state Potts model, {\it i.e.} the Ising model,
interfaces annihilate upon collisions.  The difference lies in interface
dynamics -- in the Potts models domain walls usually undergo random
walk rather than  ballistic motion.

In summary, we have analyzed a model of ballistic aggregation of
particles where each particle is approaching its nearest neighbor with
constant and universal velocity. It can be considered as an idealized
model of a world in which all neighboring countries are engaged in
permanent warfare with larger countries advancing and smaller receding.
We found a universal linear in time scaling for all types of domains and
developed a mean-field like approach to the domain size distribution
function.

\bigskip 
We are thankful to L.~Frachebourg and S.~Redner for helpful discussions.
This work was partially supported by grants from ARO and NSF.

\end{multicols}
\end{document}